\documentclass[a4paper,11pt]{article}

\usepackage{geometry}
\usepackage{amsmath}
\usepackage{amssymb}
\usepackage{graphicx}
\usepackage{hyperref}
\usepackage{enumitem}
\usepackage{cite}
\usepackage{float} 

\title{\textbf{Publicly Understandable Electronic Voting: \\ A Non-Cryptographic, End-to-End Verifiable Scheme}}
\author{Alon Gat \\ \texttt{alongat16@gmail.com}}
\date{\today}

\begin{document}

\maketitle

\begin{abstract}
Modern democracies face an existential crisis of waning public trust in election results. While End-to-End Verifiable (E2E-V) voting systems promise mathematically secure elections, their reliance on complex cryptography creates a ``black box'' that forces blind trust in opaque software or external experts, ultimately failing to build genuine public confidence.
To solve this, we introduce the concept of \textbf{Software-Free Verification (SFV)} --- a standard requiring that voters can independently verify election integrity without relying on any software. We propose a practical, non-cryptographic in-booth voting scheme that achieves SFV for national-scale elections. Our approach leverages a public bulletin board of randomized \textit{(Pseudonym, Candidate)} pairs, where a mechanically generated pseudonym is hidden among real decoy votes on a physical receipt.

Our scheme empowers citizens to audit the election using only basic arithmetic via a hierarchical Public Ledger, while anchoring the overall digital tally to physical evidence and Risk-Limiting Audits (RLAs) to guarantee systemic integrity. The result is a system that bridges the gap between mathematical security and public transparency, offering a viable blueprint for restoring trust in democratic infrastructure.
\end{abstract}

\section{Introduction}

\subsection{The Promise and Perils of Electronic Voting}
Electronic voting has the potential to offer substantial benefits over traditional paper ballots. Operationally, it provides immediate election results, reduces the manpower required for tallying, and improves overall accuracy by eliminating manual counting errors and localized fraud. Furthermore, digital interfaces greatly enhance accessibility and enfranchisement: visually impaired voters can vote independently using audio guides or enlarged fonts, illiterate citizens can identify candidates via images, and ballots can be instantly translated into multiple languages.

Despite these advantages, many cybersecurity experts warn against fully electronic voting \cite[p.~92]{nasem2018}. They argue that while physical ballot fraud does occur, its scope is inherently limited by the logistics required to execute it at scale. Conversely, because electronic voting machines rely on opaque code, a single software vulnerability or malicious exploit could silently and undetectably alter the digital tally on a nationwide scale.

\subsection{Software Independence and End-to-End Verification}
In the academic literature on election security, Rivest and Wack \cite{rivest2008} define a system as Software Independent (SI) if ``an undetected change or error in its software cannot cause an undetectable change or error in an election outcome.'' One way to achieve this is End-to-End Verification (E2E-V), allowing voters to verify that their selections were correctly recorded and that every recorded vote is correctly included in the tally.\footnote{It is important to note that E2E-V does not address the question of voter identification and eligibility, which is another key issue when it comes to public trust.}

Since E2E-V schemes often involve providing voters with receipts that contain the relevant information for verification, the E2E-V properties are typically divided into three distinct phases:
\begin{enumerate}
    \item \textbf{Cast As Intended:} The data on the receipt correctly encodes the voter's intended selection.
    \item \textbf{Recorded As Cast:} The cast ballot is accurately recorded on the public bulletin board without alteration.
    \item \textbf{Tallied As Recorded:} The tally of all recorded votes is calculated correctly and matches the official election results.
\end{enumerate}

Many of these E2E-V schemes are also \textit{receipt-free} \cite{benaloh1994} (meaning that voters are unable to prove to a third party how they voted, i.e., in any scheme involving a voter receipt, the chosen candidate cannot be inferred from the information on the receipt alone), which poses a challenge with respect to the \textit{Cast-as-Intended} phase: convincing the voters themselves that their receipt actually matches their selection. Several solutions have been proposed to reconcile these seemingly contradictory properties \cite{ronne2020, cortier2019}. For instance, the Benaloh Challenge allows voters to decide whether they want to trust the system-issued receipt and cast their ballot or ``challenge'' the system to reveal the cryptographic keys proving the connection between the receipt and their selection; if a challenge is issued, the ballot is spoiled, and the voter must repeat the process to actually cast a vote \cite{benaloh2006}.

These challenges often result in E2E-V implementations being highly complicated, relying heavily on cryptographic primitives, and impossible to verify without specialized software. Furthermore, the \textit{average voter} will almost certainly not be able to understand the verification mechanism, which in turn might undermine trust in the official election results.

\subsection{The Problem of Transparency}
In 2009, the German Federal Constitutional Court issued a landmark ruling (2 BvC 3/07) declaring that the use of electronic voting machines is unconstitutional unless ``it is possible for the citizen to check the essential steps in the election act... reliably and without special expert knowledge'' \cite{german2009}.

This legal mandate effectively disqualifies electronic voting systems that rely on complex cryptography, such as Zero-Knowledge Proofs or Homomorphic Encryption. In such systems, a voter cannot personally verify the mathematical proof; they must run a software tool to verify it for them.

Empirical research confirms that this complexity is a significant barrier to public trust. A recent study by Haney et al. found that E2E-V systems often introduce ``another layer of complexity that may be difficult for voters to understand,'' adding to an already ``complicated election ecosystem.'' Experts expressed concern that because the underpinnings of this technology include ``a lot of mathematics that most people are not going to understand,'' the systems may appear to the public as a mysterious ``black box'' \cite{haney2026}.

Furthermore, the study warns that a ``checking technology with technology'' scenario -- where voters are told to ``just trust the math'' -- is ``not acceptable'' for building genuine confidence. This complexity often leads to a ``mismatch between [voter] expectations and reality.'' As one expert noted, when a voter seeks transparency but is met with a ``string of codes'' rather than seeing their ballot in the box, they may perceive the system as ``dangerous'' because it remains primarily digital \cite{haney2026}.

Addressing these concerns requires stripping away the need to trust \textit{any} operational component whose internal workings the average person cannot physically see. Indeed, this necessitates removing blind trust not only from the central election servers and in-booth voting machines, but crucially, from the voters' personal digital devices. To truly allow the \textit{voter} to be the auditor, the integrity of the election must remain publicly verifiable even under the assumption that the verification software itself is actively malicious and colluding with the attacker.

\subsection{Defining Software-Free Verification}
To bridge the gap between the power of E2E-V and the demand for absolute transparency and public trust, we introduce a new property: \textbf{Software-Free Verification (SFV)}.

 \textbf{Definition:} A voting scheme achieves Software-Free Verification if there exists a process to verify its results in a reasonable amount of time that will remain sound even if all software used by the verifier is adversarial.
 Specifically, the scheme must allow a verifier to independently confirm that:
\begin{enumerate}
    \item Their individual vote was cast and recorded as intended.
    \item The final tally is the correct sum of all recorded votes.
\end{enumerate}

\section{Design Goals}
We aim to design a voting scheme that satisfies our SFV property, while still preserving voter anonymity and a reasonable level of coercion resistance.
We explicitly state which properties our scheme must have, and where a ``best effort'' approach could be acceptable.

\subsection{Primary Goals}
\begin{enumerate}
\item \textbf{Integrity:} The system must ensure that the final tally accurately reflects the votes cast. A total compromise of the electronic system must be detectable by the voters themselves.
\item \textbf{Dispute Resolution:} Voters must be provided with hard evidence to legally dispute a fraudulent digital tally. The system must allow the election authority and the general public to clearly differentiate between false complaints or voters misremembering their choices, and valid disputes stemming from system error or election fraud.
    \item \textbf{Private-Receipt Anonymity:} As long as the voter keeps their physical receipt private, no entity (including the election authority) can link the voter to their vote.
    \item \textbf{Exposed-Receipt Anonymity:} If a third party sees the voter's physical receipt, the voter's choice remains anonymous, provided that the observer is not colluding with the election authority or gained access to private backend system data, and does not have access to a massive, nationwide collection of private voter receipts.
    \item \textbf{Availability:} The election process must not be halted by cyber-attacks or hardware failures. Even if the entire electronic infrastructure malfunctions or is actively attacked on election day, the system must seamlessly revert to manual voting without requiring a revote or redoing the election day.
    \item \textbf{Transparency and Simplicity:} The election process must be simple, with an intuitive verification mechanism that is both easy to understand and likely to be performed by the average voter.
\end{enumerate}

\subsection{Secondary Goal}\label{sec:secondary_goal}
\begin{enumerate}
    \item \textbf{Coercion Resistance:} We prioritize mitigating simple, interpersonal coercion attempts, such as demanding to see the voter's receipt, effectively reducing this threat to the \textbf{Exposed-Receipt Anonymity} goal defined previously.
    
    \textbf{Rationale:} We accept imperfect coercion resistance as a necessary trade-off to secure both the foundational \textbf{Integrity} of the tally and the \textbf{Simplicity} of the system. We justify this compromise based on two key observations:
    \begin{itemize}
        \item \textit{Visibility of Large-Scale Attacks:} A massive, coordinated vote-buying operation inherently becomes public knowledge, allowing authorities to assess the threat and take countermeasures (up to and including reverting to manual voting). In contrast, localized coercion occurs in the shadows. Even the mere claim that systemic coercion exists (such as allegations that voters in traditional communities are pressured to mirror the choices of family or community leaders) can severely damage the perceived fairness of the election.
        \item \textit{The Illusion of Perfect Coercion Resistance:} Absolute coercion resistance is already structurally compromised in modern democracies. Remote voting methods (like mail-in ballots) and the ubiquity of smartphones allow voters to easily prove their choice to third parties.\footnote{For instance, a former Israeli minister shared a video on social media of her husband in the polling booth, accompanied by a caption stating she had required him to film the entire voting process as proof of his ballot choice. See: \url{https://web.archive.org/web/20260315134513/https://www.facebook.com/galit.distalatbaryan/videos/515254393797241/?mibextid=wwXIfr&rdid=YU2br3tKmMPdYWK4}. Such incidents demonstrate that strict coercion resistance is routinely bypassed and socially tolerated in existing physical systems.}
    \end{itemize}
\end{enumerate}

\section{The Basic Idea: A Public Bulletin Board}
At the heart of our proposed scheme is a radically simple concept: a \textit{Public Bulletin Board} of pairs $(r, C)$, where $r$ is a random pseudonym (e.g., a number) and $C$ is a candidate. If every voter can find their specific pair on this board, and the total count of pairs matches the election result, integrity is assured. This simple approach has been proposed before and largely dismissed in favor of cryptographic solutions. The main objections \cite{benaloh2015} to this simple scheme are that:
\begin{enumerate}
    \item Some voters might accidentally or deliberately reveal their pseudonyms, compromising anonymity.
    \item By being forced to reveal their pseudonyms, voters could be susceptible to coercion.
    \item \label{item:system_cheat}The system might cheat, for instance by assigning the same pseudonym to voters who are likely to vote the same way.
\end{enumerate}

Our scheme attempts to address these concerns. We argue that the first two issues could be resolved in a satisfactory fashion by hiding the voter’s $(r, C)$ pair among other valid pairs (decoys) of previous voters on a physical receipt, which removes the need for the voter to manually record their generated pseudonym. This both simplifies the voting process and ensures that even voters who would not bother to write down their pseudonym while inside the voting booth would retain the ability to participate in the verification process later.
The third issue could be partially addressed by generating the pseudonym via a \textit{Mechanical RNG} (e.g., a lottery machine). The mechanical process is intuitively understood by the layperson and is physically immune to malware, establishing a ``Physical Root of Trust'' that digital systems cannot replicate.

\section{Related Work: Comparison With Bingo Voting}
Our scheme shares significant design parallels with the \textit{Bingo Voting} scheme proposed by Bohli, Müller-Quade, and Röhrich in 2007 \cite{bohli2007}, which was the first to introduce the concept of using a mechanical random number generator for elections. While we share the core idea of a receipt where the voter's ``fresh'' random pseudonym is hidden among other pseudonyms, our implementation diverges fundamentally by using \textit{live, previously cast} votes from other citizens as decoys, while Bingo Voting hides the voter's choice through cryptographically generated decoys.

\subsection{The Problem of Collisions}
In Bingo Voting, a large number of dummy votes, pairs of $(r, C)$, are generated and cryptographically pre-committed in advance, while some of those dummy votes later appear on the voters' receipts in an attempt to make the real votes indistinguishable from the dummy votes to a potential coercer. Ultimately, the security of the scheme depends on the assumption of a trusted RNG, as the architecture is vulnerable to collisions. If the voter's $(r, C)$ pair matches a pre-committed dummy vote pair (whether due to a small number space or a rigged RNG), the system can cheat.

\subsection{Timing Trade-offs: RNG First vs. Vote First}
Crucially, in Bingo Voting, the voter selects the candidate \textit{before} the RNG draw, which prevents coercion (if the candidate were selected after the RNG draw, a coercer could issue instructions based on the random number's properties, e.g., ``Vote A if Even, B if Odd''), but compromises integrity, as the machine knows if it is ``safe'' to cheat. If the RNG outputs a dummy value associated with Candidate A, and the voter indeed chose Candidate A, the machine can simply record this transaction as the dummy vote, effectively discarding the real vote and inserting a fake one elsewhere. While sequential versions of Bingo Voting introduced hash chains \cite{henrich2012} to further mitigate this by forcing the use of specific dummies, the fundamental vulnerability of the ``Vote First'' logic remains if the RNG is compromised.

In comparison, our scheme generates the pseudonym first to prioritize integrity, further mitigating the concern raised in Objection~\ref{item:system_cheat}.

\subsection{The ``Mathematical Proof'' Privacy Flaw}
Since the dummy pseudonyms in Bingo Voting are cryptographically committed, if the Election Authority (or a hacker) leaks the opening keys (the cryptographic ``reveals'') for the used dummy votes, the anonymity of the voter is destroyed \textbf{with respect to anyone who has seen the receipt}.

The leaked keys act as a \textbf{cryptographic proof} that specific pseudonyms on the receipt are dummies. A person who saw a receipt can use the leaked data to mathematically prove which candidate the voter selected.

In contrast, our scheme avoids cryptographic commitments in favor of plaintext decoys. Even if an attacker leaks a database linking receipts to votes, they cannot \textit{prove} the link is genuine. Voters maintain \textit{Plausible Deniability}, as they can credibly claim the attacker fabricated or modified the data to sow chaos.

\section{The Protocol (Core Scheme)}

\subsection{Phase 0: Setup, Seeding, and Loading}

\subsubsection{Physical Ballot Loading}
In the morning, the polling station committee members load the voting machines with stacks of standard, pre-printed paper ballots containing checkboxes for all candidates. These are identical to manual ballots. The machine is designed to pull an anonymous ballot from the stack and mark the chosen candidate's checkbox during the voting process.

\subsubsection{Seeding the Decoy Pool}\label{sec:seeding}
To ensure that the first voters of the day have valid decoys printed on their receipts to hide their choice, polling station committee members cast one vote for every candidate before the polls officially open.\footnote{This process must occur at a minimum of one polling station per Cluster (as defined in Section \ref{sec:phase3}) to populate the shared pool. However, executing this seeding at every polling station guarantees that every machine is locally loaded with at least one valid decoy per candidate, ensuring system availability even in the event of a network failure.} These dummy votes ensure that the database is immediately populated with valid $(r, C)$ pairs for every candidate, and are mathematically deducted from the total sum when the final results are tallied. In order to guarantee that the dummy votes were not altered by a malicious system, the polling station committee members are responsible for verifying that these specific votes were included in the final Public Ledger using the standard verification mechanism described in Section~\ref{sec:phase5}.

\subsection{Phase 1: Voter Identification and Authentication}
Upon arriving at the polling station, the voter is identified by the polling station committee members using standard identification protocols. The voter's name is then crossed off a physical, printed list of eligible voters for that precinct. After successful identification, a polling station committee member manually activates a single voting session on the machine for the voter.

\subsection{Phase 2: Mechanical Randomness Generation}\label{sec:phase2}
The voter enters the booth. A transparent mechanical device physically mixes and draws a random pseudonym $r$ which remains visible for the duration of the voting session. The pseudonym $r$ is then scanned by the voting machine.

\subsection{Phase 3: User Choice and Cluster-Based Decoy Selection}\label{sec:phase3}
The voter selects Candidate $C_{choice}$ on the touchscreen. The machine must now generate a receipt. To balance coercion obfuscation with the ability to publish manually verifiable ledgers, geographically distant polling stations are grouped into pre-defined \textbf{Clusters}, which serve as the basic units for verification.\footnote{We assume coercion mostly occurs within local communities, whereas large-scale receipt gathering would be handled by authorities. Using geographically distant precincts minimizes the chance a local coercer will encounter two receipts with the same pseudonym.} \footnote{The election authority retains the ability to publish aggregate vote totals for individual precincts for general reporting; however, the Cluster remains the atomic unit for individual vote verification to preserve the anonymity provided by the cross-precinct decoy mechanism.} Previously recorded votes from within the Cluster are chosen as decoys.

\subsection{Phase 4: In-Booth Verification \& Casting}
The machine produces two physical outputs:
\begin{itemize}
    \item \textbf{Output 1 - The Marked Ballot:} The machine marks the checkbox of $C_{choice}$ on the pre-loaded ballot.
    \item \textbf{Output 2 - The Receipt:} Printed separately on a high-security, difficult-to-forge paper. The physical layout consists of $n$ rows, where $n$ is the total number of candidates. Each row displays a pair $(r, C)$ and the rows are ordered consistently across all receipts according to the candidates. The true random pseudonym $r$ appears next to the voter's selection, while the decoy pseudonyms from the wider geographic Cluster appear next to the other candidates.
    \item \textbf{Digital Signature Validation:} Every receipt includes an asymmetric digital signature generated by the machine, which cryptographically signs the complete contents of the receipt (i.e., the specific set of all $n$ $(r, C)$ pairs). The voter may choose to scan the receipt with a smartphone app that validates the signature against the specific voting machine's public key. Crucially, while this verification involves cryptography, it is an \textit{optional} stage intended only to reduce the chance that a valid claim is dismissed during a dispute.
    \item \textbf{In-Booth Verification \& Challenge Protocol:} The voter verifies that the \textbf{marked checkbox} on the ballot matches their choice and that the printed $r$ on the receipt next to their candidate of choice directly matches the $r$ still displayed by the physical mechanical generator. If the voter notices a mismatch (e.g., the physical ballot is marked for the wrong candidate, or the true $r$ appears next to the wrong candidate on the receipt), they can immediately challenge the machine. The current voting session is nullified, and the voter is allowed to vote again, optionally in the presence of a polling station committee member who can observe the session to verify the alleged machine fraud.
    \item \textbf{Casting:} The voter folds the anonymous marked ballot and inserts it into a sealed box.
\end{itemize}

\begin{figure}[H]
    \centering
    \includegraphics[width=0.67\linewidth]{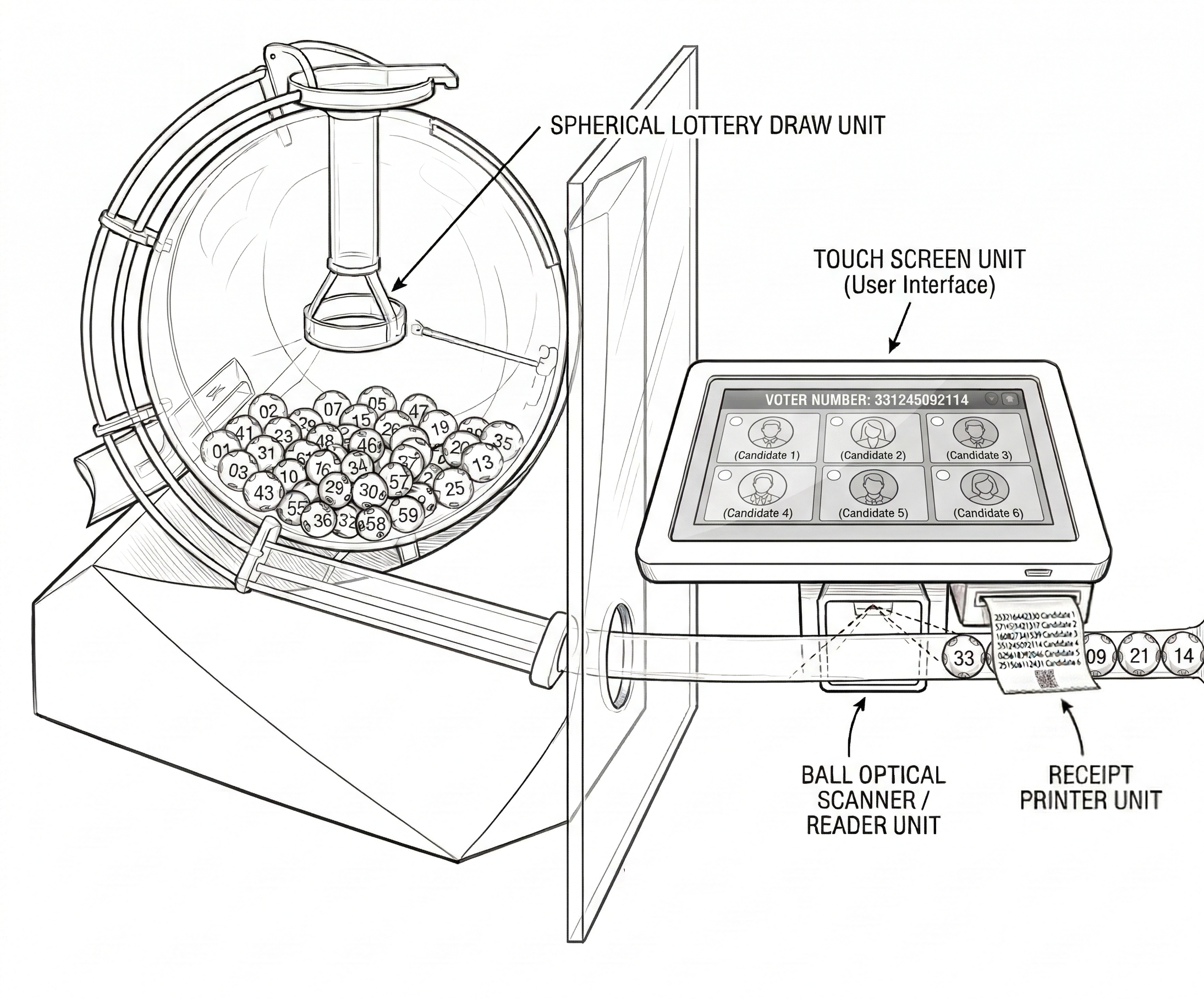}
    \caption{Lottery balls are scanned by the voting machine. The random number determines the voter's pseudonym.}
    \label{fig:placeholder}
\end{figure}

\subsection{Phase 5: Publishing Results \& Verification}\label{sec:phase5}
After the polls officially close, the election authority publishes the complete ``Public Ledger'' containing all cast votes. To support both convenient digital auditing and rigorous Software-Free Verification (SFV), the data is released in two complementary formats, and a mandatory physical audit is performed.

\subsubsection{Format A: The National Digital Ledger}
For observers willing to use digital tools, the authority releases a single, flat digital file containing all $N$ votes cast nationwide.
\begin{itemize}
    \item \textbf{Content:} A list of all $(r, C)$ pairs from every precinct in the country.
    \item \textbf{Sorting Logic:} To facilitate search, the entire list is sorted strictly by the random pseudonym $r$.
    \item \textbf{Digital Verification Method:} Voters can locate their specific $(r, C)$ pair instantly using standard digital search functions, or by simply scrolling through the file. The total number of rows in the file corresponds to the total national vote count plus the predefined number of dummy votes cast during the seeding phase. Consequently, the actual number of votes each candidate received can be easily determined digitally by counting the candidate's total occurrences in the file and subtracting their designated number of dummy votes.
\end{itemize}

\subsubsection{Format B: The Hierarchical Ledger (For Manual SFV)}
To enable manual, software-free summation where a human cannot process millions of rows, the authority also publishes the data formatted using a \textbf{Hierarchical Cluster Architecture}.

The ledger is partitioned logically into \textbf{Clusters}, which are aggregations of several precincts. The number of votes within a Cluster must be small enough that a manual pass over the printed data is feasible for a human (for example, around 5,000 votes).\footnote{Cluster size should be chosen as a trade-off between SFV and coercion-resistance. If a manual verification process is deemed unnecessary, the Cluster can encompass the entire country.}

The ledger hierarchy is structured as follows:
\begin{itemize}
    \item \textbf{Clusters:} The atomic unit for manual verification. The votes within a Cluster are published as a distinct list. To support manual counting, this list is grouped by Candidate, then sorted by $r$, with continuous serial numbers applied to the final sorted order.
    \item \textbf{Super-Clusters:} An aggregation of base Clusters (for example, 10 Clusters in a Super-Cluster).
    \item \textbf{Ultra-Clusters:} An aggregation of Super-Clusters (for example, 10 Super-Clusters in an Ultra-Cluster), accommodating additional hierarchical levels as dictated by the size of the electorate.
\end{itemize}

\textbf{Manual Verification Process (SFV):}
Voters seeking Software-Free Verification can achieve it in a reasonable amount of time using only basic addition and subtraction:
\begin{enumerate}
    \item \textbf{Top-Level Verification:} The voter sums the official totals of the highest aggregated levels (e.g., Ultra-Clusters) to ensure they match the national total.
    \item \textbf{Drill-Down Verification:} The voter selects their specific hierarchical branch, summing its immediate sub-tiers to ensure they match the reported total. They repeat this process down the chain until they reach their specific base Cluster.
    \item \textbf{Base Tally \& Individual Verification:} The voter prints the raw data file for their specific base Cluster. Because the list is grouped by candidate and numbered sequentially, the voter simply verifies that the serial numbers increment perfectly without gaps from 1 to $N_{Cluster}$. They can then instantly calculate the total votes for each candidate: $(Last\ Serial\ Number) - (First\ Serial\ Number) + 1 - (Candidate\ Dummy\ Votes\ In\ Cluster)$. Finally, they easily locate their specific $(r, C)$ pair within their chosen candidate's block to confirm their vote was perfectly recorded.
\end{enumerate}

\subsubsection{Risk-Limiting Audit (RLA) of Physical Ballots}\label{sec:rla}
To provide a systemic layer of integrity that does not rely solely on individual voter initiative, the election authority shall perform a mandatory Risk-Limiting Audit (RLA). Immediately following the close of polls and before the final results are certified, a rigorous statistical sample of physical ballot boxes is selected for a manual count in the presence of representatives of all registered competing parties. Using established RLA methodologies, this process ensures that if the reported electronic outcome is incorrect, there is a large, pre-specified probability (the ``risk limit'') that the audit will detect it. If the discrepancy between the electronic result and the manual sample exceeds the threshold defined by the risk limit (indicating that the electronic tally is statistically inconsistent with the physical evidence), a full manual count of all ballots in the affected region is automatically triggered.

\subsection{Phase 6: Dispute Resolution}
If a voter successfully completes Phase 5, their trust in the tally is verified. However, if voters discover a discrepancy (e.g., their specific $r$ is missing or associated with an incorrect candidate), the scheme provides a mechanism to contest the result:
\begin{enumerate}
    \item \textbf{Evidence Presentation (Physical Layer):} Voters present their physical, printed receipts to the election tribunal or designated committee. Crucially, the primary evidence of systemic fraud is physical and sociological: the mere existence of a large volume of genuine-looking, hard-to-forge receipts presented by many unrelated, ordinary citizens is in itself a massive indicator of a compromised election.
    \item \textbf{Cryptographic Authentication (Supplementary Layer):} To further aid the investigation, the designated committee can optionally use the machine's public key to verify the asymmetric digital signatures on the contested receipts. If the signature is verified and there is a mismatch between the receipt and the results, there is indisputable evidence that the machine or the system has been compromised, so a manual counting of the ballots is triggered at least for that precinct.
    \item \textbf{Recount Trigger:} A manual tally of the physical ballots is performed if the election committee determines that the reported disputes might be credible.
\end{enumerate}

\subsection{Security Mechanisms and Practical Considerations}
While the core protocol outlines the primary voting flow, several defensive mechanisms and design choices are embedded within the system to ensure resilience against specific attack vectors:

\begin{itemize}
    \item \textbf{Hardware Fallbacks and DoS Protection:} If the voting machines fail or are attacked, the identical, pre-loaded paper ballots are removed from the tray and used for immediate manual voting. To prevent a compromised machine from rapidly destroying this paper supply (a Denial of Service attack on the physical fallback), the hardware physically restricts the mechanism to pull and print only one ballot at a time, making a distinct, audible noise during the process.
    \item \textbf{Anti-Stuffing Defense:} The physical voter roll serves as an independent count of the total number of participants. At the end of the election, the number of crossed-off names is compared to the total number of votes recorded by the machine (taking into account the dummy votes cast for each candidate before the polls open). This ensures that a compromised machine cannot ``stuff the ballot box'' with non-existent votes. If a discrepancy is detected between the number of signed voters and the electronic tally, the sealed physical ballots are manually counted to resolve the issue.
    \item \textbf{Entropy and Usability Trade-off:} The distribution of $r$ does not have to be perfectly uniform, provided there is sufficient entropy. The length of $r$ represents a trade-off: the sample space must be large enough to minimize expected collisions, yet short enough for a voter to easily verify visually.
    \item \textbf{Receipt Shredder:} To prevent potential coercers from harvesting a large number of discarded receipts, we recommend including a designated in-booth paper shredder.
    \item \textbf{UI Randomization for Side-Channel Defense:} The list of candidates could be displayed in a randomized order for each individual voter on the touchscreen. This mitigates physical side-channel attacks, such as a coercer or subsequent voter deducing the previous voter's choice based on fingerprint smudges left on the screen.
\end{itemize}

\section{Security Analysis: A Defense-in-Depth Approach}
Instead of analyzing the system under a single, static threat assumption, we evaluate its security and privacy properties across an escalating spectrum of potential compromise, as well as against external coercion. The fundamental security of our Software-Free Verification (SFV) scheme relies on the probabilistic guarantee that an adversary cannot selectively alter a significant percentage of votes without triggering a mathematically proportional risk of detection by independent, offline verifiers.

\subsection{Scenario 1: Foreign Influence - Vote Buying Operation}
\textbf{Assumption:} The internal election apparatus (voting machines, central servers, and RNG modules) operates honestly, but a hostile foreign or domestic entity attempts to subvert the results through economic means.

\textbf{Automated Vote-Buying:} A sophisticated coercer might attempt a large-scale vote-buying operation, providing a cryptocurrency reward to the first person who submits a valid random pseudonym that turns out to correspond to a vote for the candidate the coercer is trying to promote. The usage of \textit{fresh} decoys from within the wider Cluster would somewhat mitigate such an operation, because the coerced voter would be forced to race against other people who receive that exact same random pseudonym on their receipts roughly at the same time. Furthermore, as discussed in Section~\ref{sec:secondary_goal}, any large-scale vote-buying operation would inherently become public knowledge. This visibility gives authorities the opportunity to assess the situation and implement escalating countermeasures. Initial steps could include instructing polling station committee members to strictly enforce bans on smartphones inside the voting booths, preventing the coerced voter from transmitting the pseudonym in real-time and forcing them into the timing race. In more extreme cases where the threat to election integrity is deemed severe, authorities can take the decisive action of reverting to manual voting.

\subsection{Scenario 2: Election Officials Malfeasance - Ballot Stuffing and Coercion}
\textbf{Assumption:} The internal election apparatus operates honestly, but polling station committee members abuse their administrative access to illegitimately cast votes using the identities of absent voters, or attempt to leverage their insider knowledge to coerce voters.

\textbf{Hardware Bottlenecks on Ballot Stuffing:} In a purely paper-based election, corrupt officials can rapidly stuff a ballot box to inflate candidate totals. In our electronic scheme, this fraudulent activity is actively bottlenecked by the system's physical hardware. Because the machine requires the mechanical RNG to physically draw a pseudonym, and the printer must generate a unique receipt and marked ballot for every transaction, insiders cannot instantaneously inject batches of digital votes. Furthermore, anomalous voting rates or an irregular seeding phase would be easily detectable by inspecting the system's logs.

\textbf{Coercion of Early Voters:} Because polling station committee members cast the initial dummy votes, they know the specific pseudonyms entering the decoy pool. Theoretically, they could attempt to coerce the first few voters of the day, capitalizing on the higher probability that their specific dummy votes will appear on those early receipts. However, this risk is partially mitigated if multiple dummy votes are cast per candidate across the broader Cluster, significantly diluting the predictability of the initial decoy pool.

\subsection{Scenario 3: Limited Cyber Attack - Network Denial of Service}
\textbf{Assumption:} The attacker cannot breach the voting machines' internal software or the RNG, but they successfully execute a Denial of Service (DoS) attack that severs communication between localized voting machines and the wider network.

\textbf{Impact on Decoys and Privacy:} If machines are isolated from the network, they can no longer fetch fresh decoys from the broader Cluster. To maintain availability, the system must fall back to using its localized history of previous voters to generate decoys. While this ensures the election is not halted, it explicitly degrades exposed-receipt anonymity and coercion resistance. A local coercer could potentially map the limited pool of localized random pseudonyms to voters in their immediate vicinity. For instance, if a coercer observes the same decoy pseudonym associated with a specific candidate across multiple receipts, they can mathematically deduce that at most one of those voters actually selected that candidate, thereby weakening the plausible deniability of the decoy pool.

\subsection{Scenario 4: Malicious USB Attack - Partial Software Compromise}
\textbf{Assumption:} An attacker physically compromises a limited number of voting machines within specific polling stations (e.g., via an infected USB stick or insider access), allowing them to act maliciously without controlling the central infrastructure or the entire network.

\textbf{Impact on Decoys:} A compromised machine can intentionally lie to honest machines within its Cluster by providing predictable, correlated, or malicious decoy numbers. If a voter's receipt is seen by a coercer, this malicious decoy data can degrade the exposed-receipt anonymity and coercion resistance for voters in that localized Cluster.

\textbf{Damage Containment via Local Recounts:} If the malicious machine attempts to fake the actual election results by dropping or altering votes before sending them to the central server, it is constrained by the verification loop. This localized fraudulent behavior may be detected by voters from the affected precinct verifying their votes. Once voters from the affected area report the mismatch and present their physical receipts, a manual recount of the sealed ballot boxes is triggered \textit{only} for those specific precincts. The corrupted electronic tally is discarded and replaced by the authoritative physical count, effectively limiting the damage to the local level and preserving the integrity of the wider election without requiring a nationwide recount.

\textbf{Availability (Localized DoS):} If the compromised machines resort to a blunt Denial of Service attack by crashing or printing garbage, the local polling station committee members simply move the identical, un-ruined pre-loaded paper ballots to a manual voting table, maintaining continuous voting for those precincts.

\subsection{Scenario 5: Comprehensive Cyber Attack - Full Software Compromise}
\textbf{Assumption:} A sophisticated cyber attack successfully compromises all voting machines and the main election backend server. However, the physical/mechanical RNG remains uncompromised and perfectly random.

\textbf{The In-Booth Verification Barrier:} Because the mechanical RNG is physically independent and transparent, the malicious software cannot force, dictate, or predict the true random pseudonym $r$. The machine is forced to print the visually verified $r$ next to the voter's chosen candidate on the receipt. If the software attempts to cheat inside the booth by printing a fake pseudonym or shifting the true $r$ to a different candidate, the voter immediately spots the discrepancy between the physically generated pseudonym and the pseudonym on the receipt, exposing the fraud before the vote is even cast.

\textbf{Verification and Nationwide Recount:} Unable to safely forge the printed receipt in the booth, the malicious software's only option to manipulate the tally is to drop or alter the valid votes later on the central server. However, the End-to-End Verifiable nature of the system ensures that voters checking the Public Ledger will notice if their votes are missing or flipped. Any large-scale attempt to change, remove, or add votes will, in all likelihood, trigger a large number of credible dispute claims backed by valid physical receipts. This would lead to a manual recount of the ballots, overriding the fraudulent electronic results entirely.

\subsection{Scenario 6: Rigged Election - Full Software and Hardware Compromise}
\textbf{Assumption:} The government is holding a sham election where the voting machines, central election server, and the physical mechanical RNG modules are all colluding to rig the election (e.g., RNGs compromised via hidden electromagnets or tampered mechanics).

\textbf{Expected Number of Collisions:}
We can model the expected collision rates for a uniformly distributed random pseudonym $r$ using the generalized Birthday Problem. Let $N$ be the total number of eligible voters and $S$ be the size of the sample space.

Using the Poisson approximation for the binomial distribution, the expected number of voters assigned to a specific random pseudonym is $\lambda = N/S$. The expected number of random pseudonyms generated exactly $k$ times across the national ledger, denoted as $E[C_k]$, is:
\[E[C_k] \approx S \cdot \frac{\lambda^k}{k!} \exp(-\lambda).\]

For pairs ($k=2$), the expected number of double-collisions across the entire ledger is:
\[E[C_2] \approx S \cdot \frac{(N/S)^2}{2} \exp\left(-\frac{N}{S}\right) = \frac{N^2}{2S} \exp\left(-\frac{N}{S}\right).\]

To demonstrate this with concrete parameters, assuming a sample space of $S = 10^{12}$ (a 12-digit number) and an electorate of $N = 10^7$ (10 million voters), we get $\lambda = 10^{-5}$. Plugging these values into our equation for pairs:
\[E[C_2] \approx 10^{12} \cdot \frac{(10^{-5})^2}{2} \exp(-10^{-5}) \approx 50.\]

Because the occurrence of natural collisions follows a Poisson distribution, the variance is approximately equal to the mean ($\sigma^2 \approx E[C_2] \approx 50$), yielding a standard deviation of $\sigma \approx 7.07$. Therefore, under these parameters, a normal election will almost certainly ($>99.7\%$ confidence) exhibit between 28 and 72 natural double-collisions across the entire national ledger.

\textbf{The Homogeneous Precinct Prediction Attack:}
Because the RNG is rigged, the attacker no longer faces the strict in-booth collision trap created by true randomness. The attacker can actively attempt to create collisions by forcing the machine to draw a random pseudonym $r$ that has already been used by a previous voter.

The success of this attack relies heavily on prediction. If the attacker targets a heavily homogeneous precinct where they can accurately predict the voter's choice (e.g., they know there is a $90\%$ chance the voter will choose Candidate A), they can attempt to steal a vote using a ``Collision Budget'':
\begin{itemize}
    \item \textbf{Successful Steal:} The attacker forces a reused pseudonym $r$ (originally associated with Candidate A within the same Cluster). The current voter indeed selects Candidate A. The machine records only a single $(r, C_A)$ pair on the Public Ledger. The machine has effectively absorbed two votes into one entry without triggering a public collision. The attacker can now safely inject a fraudulent vote for their preferred candidate elsewhere in the same Cluster without changing the total vote count.
    \item \textbf{Failed Steal (Visible Collision):} The attacker forces the reused pseudonym $r$, but the voter unexpectedly selects Candidate B. The machine cannot absorb this. It must either publish a visible collision on the ledger (associating $r$ with both A and B) or risk immediate in-booth detection (such as printing a mismatched receipt). To avoid a physical confrontation in the booth, the attacker publishes the visible collision.
\end{itemize}

\textbf{Mathematical Limits of the Collision Budget:}
To successfully alter an election result without triggering public mathematical detection, the attacker must not exceed a reasonable statistical variance for natural collisions. As calculated previously, assuming a 12-digit number uniformly distributed in a 10-million voter election, the average expected number of double-collisions is $50$, with a standard deviation of $\sigma \approx 7.07$. An attacker might introduce roughly $+20$ extra visible collisions without arousing statistical suspicion. This $+20$ acts as their strict ``Collision Budget''.

If the attacker boasts a highly accurate $90\%$ prediction rate in homogeneous precincts, they face a $10\%$ expected failure rate per attempt. To exhaust a budget of $B = 20$ visible collisions at a failure rate of $(1 - p) = 0.10$, the attacker can make an expected $200$ total attempts ($B / (1-p)$). Out of these $200$ attempts, they are expected to successfully guess correctly and steal about $180$ votes.

Stealing a few hundred votes in a national election of 10 million voters is statistically insignificant. To sway an election by hundreds of thousands of votes using this predictive method, the attacker would unavoidably generate tens of thousands of failed attempts. The machine must either publish these as visible collisions on the Public Ledger -- creating an astronomical spike easily identified by any citizen auditing the file -- or risk immediate in-booth detection at a massive scale by printing mismatched receipts. Either outcome provides undeniable proof of systemic fraud.

\subsection{Scenario 7: Doomsday - Full Software, Hardware, and Verification Devices Compromise}
\textbf{Assumption:} The adversary controls all elements in Scenario 6, \textit{and} has successfully infected the personal verification devices (smartphones, PCs) of \textit{all} citizens. The verification software itself is actively colluding with the central election server, giving the attacker full control over what is displayed to any individual user.

\textbf{The Statistical Audit Constraint:}
It is important to note that the attacker is constrained by the mandatory Risk-Limiting Audit (RLA) of physical ballots (Section~\ref{sec:rla}), preventing them from rigging the results by an amount large enough to be detected by the physical sampling of ballot boxes.

\textbf{The Limits of Offline Communication:}
To accurately model the attacker's advantage, we assume a worst-case scenario regarding offline defense: verifiers do not exchange raw ledger files offline. Our model assumes that the only piece of information citizens can reliably and safely exchange offline, which is entirely immune to the attacker's digital manipulation, is the final, top-level election tally broadcast on public television or radio. Under this assumption of a lack of granular offline communication, the attacker could theoretically show one fake version of the detailed ledger to one verifier and a completely different fake version to another verifier without them immediately realizing the discrepancy.

\textbf{Digital Verification \& Information Gathering:}
Even with full control over the user's interface, successfully fooling the verifier is not easy if the attacker does not know the connection between the verifier and their specific vote in advance. However, the attacker can harvest information to isolate specific voters. By monitoring compromised devices, the malware can track specific random pseudonyms typed into search queries, observe where users pause while scrolling, or exploit optional election-day features like users scanning the digital signatures on their receipts.

Through these methods, the attacker can try to gather specific ``votes'' (random pseudonyms paired with their candidates) that belong to voters they believe are highly unlikely to engage in manual verification. We model this by assuming the attacker has successfully compiled a certain number of ``safe votes'' that they know they can safely alter with a very low risk of provoking public complaints. Nonetheless, given that anyone can serve as a verifier without expert knowledge, executing this flawlessly at scale is immensely difficult. Even a small number of fraud claims deemed credible by the public will rapidly trigger a snowball effect, generating widespread public scrutiny and a surge in verification attempts.

\textbf{The Manual Offline Verification Defense:}
To evaluate the system's resilience, we model a scenario where a subset of citizens perform the manual verification process (as defined in Section~\ref{sec:phase5}) without relying on their own compromised personal devices -- for instance, by downloading and printing the Cluster file from a public or independent computer located in a different city from their primary residence. This geographic displacement prevents the attacker from using location data to optimize their fraud. If an attacker observes a download request originating from a specific city, they might strategically classify that area as ``monitored'' and choose to leave those votes untouched, instead altering votes in distant ``safe precincts'' to avoid detection. By verifying the ledger from a location far removed from where they cast their ballot, the voter effectively denies the attacker this safe haven strategy.

\textbf{Probability of Evading Detection (A Tight Race Model):}
To estimate the attacker's chances of getting away with systemic fraud against careful manual verifiers, we model a tight national race where Candidate A legitimately receives $51\%$ of the vote and Candidate B receives $49\%$, and the attacker wants to flip the result.
\begin{itemize}
    \item Let $N = 10,000,000$ be the total number of votes.
    \item Let $A = 5,100,000$ be the number of ``actionable'' votes (the total votes originally cast for the candidate the attacker wishes to steal from).
    \item Let $M = 200,000$ be the total number of votes the attacker must flip.
    \item Let $K$ be the pool of actionable votes the attacker has identified as ``safe'' to alter. We model an attacker with $K = 100,000$.
    \item Let $V_{\text{manual}}$ be the number of manual verifiers in our model whose vote identity remains unknown to the attacker.
\end{itemize}

To present a mathematically coherent printed ledger, the attacker must alter exactly $M = 200,000$ votes. They safely alter all $K = 100,000$ known votes. However, because $M > K$, the attacker is mathematically forced to change the remaining $100,000$ votes completely blindly from the remaining pool of $A - K = 5,000,000$ unknown actionable votes.

Because this modeled manual verifier's specific vote is hidden somewhere within that remaining pool, the probability ($P_{\text{catch}}$) that the attacker accidentally alters the verifier's vote (thus exposing the fraud on the printed page) during a \textit{single} manual verification is:
\[P_{\text{catch}} = \frac{M - K}{A - K} = \frac{200,000 - 100,000}{5,100,000 - 100,000} = \frac{100,000}{5,000,000} = 2\%.\]

Since a single isolated fraud claim might be easily dismissed, we examine the probability that the fraud is independently caught by at least 10 manual verifiers.

Assume there are only $V_{\text{manual}} = 1,000$ careful manual verifiers nationwide (just $0.01\%$ of the voting population). The expected number of times the attacker will be caught is $\mu = 1000 \times 0.02 = 20$ catches.

Using the binomial distribution, the probability that the attacker gets caught by 10 or more independent manual verifiers is approximately \textbf{99.5\%}. This rough calculation demonstrates that even under the extreme assumption of a full compromise of all verification devices, some careful private auditors, without any deep technical knowledge or expertise in cryptography or computer engineering, are highly likely to detect a discrepancy.

\subsection{Inherent Limitations: Timing Side-Channel Attacks}
It is important to note a vulnerability inherent to \textit{all} electronic voting machines that directly record votes (DRE systems). If an attacker can cross-reference physical information (e.g., security cameras or polling station committee logs recording the exact timestamp a specific voter entered the booth) with the internal electronic timestamp logs of the voting machine, they can effectively de-anonymize the voter.

\section{Advanced Features}
Although the core scheme provides robust security for typical scenarios, specific high-threat environments may inspire advanced defenses. However, we do not necessarily recommend deploying these features. They trade simplicity for theoretical resilience, could increase implementation costs and potentially introduce new, unforeseen attack vectors onto the system as a whole. They are presented here merely as conceptual mitigations for highly sophisticated attacks.

\subsection{Advanced Feature 1: Voter-Injected Entropy for Heavily Homogeneous Precincts}\label{sec:adv_feature1}
In the basic scheme under the threat of a fully rigged authority (Scenario 6), a machine with complete physical control over the mechanical RNG (e.g., via hidden magnets) could theoretically attempt to force specific random pseudonyms. In a heavily homogeneous precinct (where the attacker can anticipate the vote with nearly absolute certainty), the machine could generate the same pseudonym $r_{base}$ for multiple voters but record only one instance in the Public Ledger, effectively deleting the real votes and replacing them with fraudulent votes for the attacker's preferred candidate.

To mitigate this, we can introduce \textbf{Voter-Injected Entropy}, allowing the voter to influence the generation of a small part of the pseudonym (after the generation of the larger part) without actually controlling it (e.g., by choosing 1 of at least 3 concealed lottery balls before revealing them to control the last digits, or by adding a chosen digit to a machine-committed digit mod 10).
\begin{itemize}
    \item \textbf{Defense Mechanism:} This creates a deadlock that secures the system against both parties. The machine cannot force a collision (vote stealing) because it cannot predict the user's entropy choice. Simultaneously, the user cannot dictate the final pseudonym (coercion resistance) because they cannot see the machine's concealed component. This ensures that a coercer cannot successfully demand a specific output pattern (e.g., ``Make the last digit 7'') to verify compliance on the receipt.
    \item \textbf{Statistical Detection:} A collision in the part of the pseudonym that is not influenced by the voter will be defined as a \textbf{semi-collision}. If the system attempts to cheat by introducing a collision on the larger part of the pseudonym (hoping that the later voter-influenced part would also match), it would risk introducing semi-collisions to the Public Ledger. Although a few semi-collisions for two pseudonyms might be dismissed as a standard Birthday Paradox occurrence ($\sim \sqrt{N}$), the occurrence of semi-collisions for three or more pseudonyms for the same candidate should be statistically negligible, so that the appearance of ``Triplets'' in the ledger becomes an indisputable, public proof of fraud.
\end{itemize}

\subsection{Advanced Feature 2: AI Coercion Assistant}
In a hypothetical scenario where AI assistants are already incorporated into voting machines (for example, to provide Natural Language Processing interfaces for increased accessibility for people with disabilities), this capability could be leveraged to further assist with coercion resistance.

The machine's existing AI interface could offer a discreet ``Coercion Help'' mode that interprets natural language requests to filter the decoy pool:

\begin{itemize}
    \item \textbf{Scenario A: Algorithmic Coercion.} Sophisticated coercers might employ conditional instructions (e.g., ``Vote for Candidate A if your number is even, and B if it is odd''). A voter wishing to vote for Candidate C can simply input this instruction into the interface. The AI parses the logic and retrieves specific decoys that satisfy the condition, constructing a receipt that validates the ``cover story'' while registering the true vote for C.
    
    \item \textbf{Scenario B: Social Context Coercion.} A voter might fear that a family member will recognize their own random pseudonym on the voter's receipt. The voter could state: ``My husband voted on this machine about 2 hours ago, so don't use decoy votes from that time.'' The AI would then filter the available decoy pool to exclude that specific time window, ensuring the husband's random pseudonym does not appear on the receipt, thereby preventing him from deducing that his wife voted for a different candidate.
\end{itemize}

Note: By incorporating this feature, we can also further simplify advanced feature 1 (Section~\ref{sec:adv_feature1}), as in the absence of the coercion threat the voter can directly control part of the pseudonym (e.g., by choosing the last digit).

\section{Conclusion}
In this paper, we have proposed a non-cryptographic End-to-End Verifiable (E2E-V) electronic voting scheme designed around the principle of Software-Free Verification (SFV). By shifting away from complex mathematics, our approach allows the average voter to intuitively understand the verification mechanism and execute it independently, showcasing that it is entirely possible to remove trust even from the underlying verification software and digital devices.

We postulate that a simple, accessible verification mechanism will result in a significantly higher percentage of the electorate actively participating in the audit process compared to what can be expected from traditional cryptography-based schemes.

Ultimately, by bridging the gap between the theoretical appeal of verifiable elections and the practical democratic demand for transparent public processes, we hope our scheme provides a viable blueprint for the responsible integration of technology into national elections, simultaneously restoring and strengthening public trust.

\end{document}